\documentclass[3p,times,procedia]{elsarticle}
\usepackage{nupha_ecrc}

\volume{00}

\firstpage{1}

\journalname{Nuclear Physics A}

\runauth{}

\jid{nupha}

\jnltitlelogo{Nuclear Physics A}

\usepackage{floatrow}
\usepackage{amssymb}

\usepackage{lineno}

\usepackage[figuresright]{rotating}

\begin{document}

\begin{frontmatter}



\dochead{XXVIIth International Conference on Ultrarelativistic Nucleus-Nucleus Collisions\\ (Quark Matter 2018)}

\title{The QCD Phase Diagram from Statistical Model Analysis}


\author[2,4]{R.Stock}
\author[1]{F.Becattini}
\author[2,3]{M.Bleicher}
\author[2]{J.Steinheimer}

\address[2]{Frankfurt Institute for Advanced Physics(FIAS), Frankfurt, Germany}
\address[4]{Institut fuer Kernphysik, Goethe-Universitaet, Frankfurt, Germany}
\address[1]{Universita di Firenze and INFN Sezione di Firenze, Firenze, Italy}
\address[3]{Institut fuer Theoretische Physik, Goethe-Universitaet, Frankfurt,Germany}

\begin{abstract}
In high multiplicity nucleus-nucleus collisions baryon-antibaryon annihilation
and regeneration occur during the final hadronic expansion phase, thus distorting
the initial equilibrium multiplicity ratios. We quantify the modifications
employing the hybrid UrQMD transport model and apply them to the grand
canonical partition functions of the Statistical Hadronization Model(SHM). We analyze
minimum bias and central Pb+Pb collision data at SPS and LHC energy. We explain the Pion to Proton ratio puzzle. We also reproduce the deuteron to proton
ratio at LHC energy by the SHM, and by UrQMD after attaching a phase space
coalescence process. We discuss the resulting ($T$,$\mu_{B}$) diagram.
\end{abstract}


\end{frontmatter}


\section{Introduction}
\label{Introduction}
The boundary between partons and hadrons in the QCD phase diagram can be studied
experimentally in relativistic nucleus-nucleus collisions. The collisional
system crosses the hadronization domain/line after a preceding phase
of hydrodynamic expansion and cooling. Several physics observables assume their
asymptotic values at, or in the close vicinity of the hadronization "point" in the
(T,$\mu_{B}$) plane (they "freeze out"), notably the multiplicities of the various hadronic
species. We note that no analytic model exists for the process of hadronization
because the transition temperature falls well into the nonperturbative domain of QCD.
One thus has to turn to QCD inspired models such as the string- or cluster-formation
pictures. The latter was formulated by Amati and Veneziano\cite{1}, to describe hadron formation in electron-positron annihilation. In it, the primordial DGLAP shower
evolution terminates with colour neutralization in local multi-parton singlet clusters which then decay quantum mechanically onto the QCD hadron/resonance spectrum, respecting invariant mass and quantum number conservation. This decay leads to hadron production according to statistical equilibrium via species-wise phase space weights.
This highly non-trivial phenomenon has found a twofold confirmation, in a
microscopic implementation in the parton-hadron transport model by Ellis and Geiger\cite{2},and in the analysis by Becattini\cite{3} of LEP hadron production multiplicities in e$^{+}$e$^{-}$ annihilation with the statistical hadronization model(SHM) which reported a
temperature of $160$~MeV. The partition functions of the SHM represent the phase space
weights, and the SHM analysis extracts from multiplicity data the temperature prevailing
at hadronization, along with the baryochemical potential expressing conservation laws:
the parameters of the QCD phase diagram.

No baryonic final state interaction can distort the multiplicity ratios in elementary
collisions. However that could be different in high multiplicity central A+A collisions. We have lately demonstrated\cite{4,5,6} that post-hadronization baryon-antibaryon annihilation is an important feature of the hadronic expansion phase, which drives the
system away from the primordially imprinted hadro-chemical equilibrium. This implies an
actual distinction between hadronization and hadrochemical freeze-out. The set of
thus-created multiplicities thus develops tensions ("non thermal" ratios), and the
deduced temperature changes with sub-selections of considered species\cite{7}. The effects of
the "afterburning" stage were taken into account by determining modification factors
for the various multiplicities by a hybrid UrQMD transport model calculation that traces
the hadron/resonance population from initial Cooper-Frye hadronization to final
decoupling. These modification factors are then employed as multipliers of the
corresponding grand canonical partition functions in the SHM model, thus reconstructing the "true" hadronization point. We consider the resulting, modified
(T,$\mu_{B}$) values as entries in the QCD phase diagram: the goal of this investigation.

We analyze sets of hadronic multiplicities from the ALICE LHC experiment, for
minimum bias and central Pb+Pb collisions at $2.76$~TeV, and from NA49 and E-802 for central Pb+Pb/Au+Au collisions at the SPS/AGS energies $17.3$, $8.7$, $7.6$ and $4.75$~GeV (see ref.~\cite{6} for references). At the AGS energy a sharp drop of the deduced temperature
occurs. At the higher energies(lower $\mu_{B}$) the curve is near flat, in
agreement with Lattice QCD predictions(~\cite{8,9}). The temperature at $\mu_{B}$ near zero is
found to be $164\pm3$ MeV. We also include the ALICE data for Deuteron production(10) in
the SHM analysis finding perfect agreement at the same temperature. Finally, in order
to test the agreement between SHM and coalescence model descriptions of such cluster
yields we have attached\cite{11} a phase space coalescence mechanism to the UrQMD model
and to its hybrid version, addressing both the ALICE p+p and Pb+Pb data for the
D/P ratio at $2.76$~TeV. This model gives an excellent description over the entire
range of midrapidity multiplicity densities.
%
\begin{figure}[t]
%
%
%
\floatbox[{\capbeside\thisfloatsetup{capbesideposition={right,bottom},capbesidewidth=6cm}}]{figure}[\FBwidth]
{\caption{UrQMD modification factors vs. centrality in Pb+Pb at 2.76TeV.}
\label{fig1}}
{\includegraphics[width=0.95\linewidth]{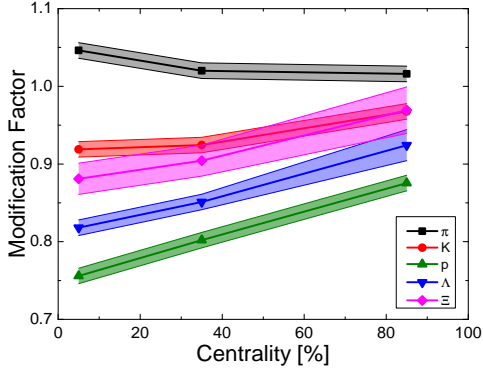}}
\end{figure}
%
\section{UrQMD afterburning and modification factors in the statistical model}
\label{urQMD}
As all details of the employed data and the model calculations have been
extensively described in previous publications\cite{5,6} we restrict here to a brief
overview. We studied the afterburner effects with the hybrid UrQMD hadron
transport model. The hydro-evolution was stopped, generally speaking, once the
energy density (or, alternatively, the temperature) falls below a pre-set
critical value in the fluid cell under consideration, then calculating the
hypersurface element with a state of the art hypersurface finder. Then the
Cooper-Frye equations were sampled in accordance with conservation of charges
and energy, and the resulting particle vector information transferred to the
cascade part of the UrQMD model. The effect of final state interaction was then
quantified by either stopping the calculation directly after hadronization,
letting the produced species undergo all their strong decays as if in vacuum,
thus establishing a fictitious multiplicity set ideally referring to the
hadronization "point". Alternatively, the final afterburner UrQMD stage was
attached, and the multiplicities at decoupling generated. For each hadronic
species we obtain the ratio of the two respective multiplicies, the so-called
modification factor. This factor got finally employed in the SHM analysis.
We show in Fig.\ref{fig1} the modification factors for pions, kaons, protons, Lambda and
Xi, obtained for Pb+Pb collisions at $2.76$~TeV, in a selected central,
semiperipheral and peripheral centrality window, corresponding to $0-10\%$,$ 30-40\%$
and $80-90\%$ cuts in the charged particle midrapidity density distribution. One
sees that the mesons remain relatively unaffected by final state effects while
the baryons (and the corresponding antibaryons) are subject to annihilation
losses. Not surprisingly the effects diminish toward peripheral collisions, due
to a shortening of the expansion duration. The seemingly small increase of the
central pion yield, of about $5\%$, does in fact correspond to an extra pion number of
about $110$, an indeed non-thermal effect which we, consequently, ascribe to  baryon-
antibaryon annihilation to pions. This explains the pion to proton ratio puzzle\cite{12} of the ALICE data, almost quantitatively.
%
\begin{figure}[t]
\begin{center}
\includegraphics[width=0.38\linewidth]{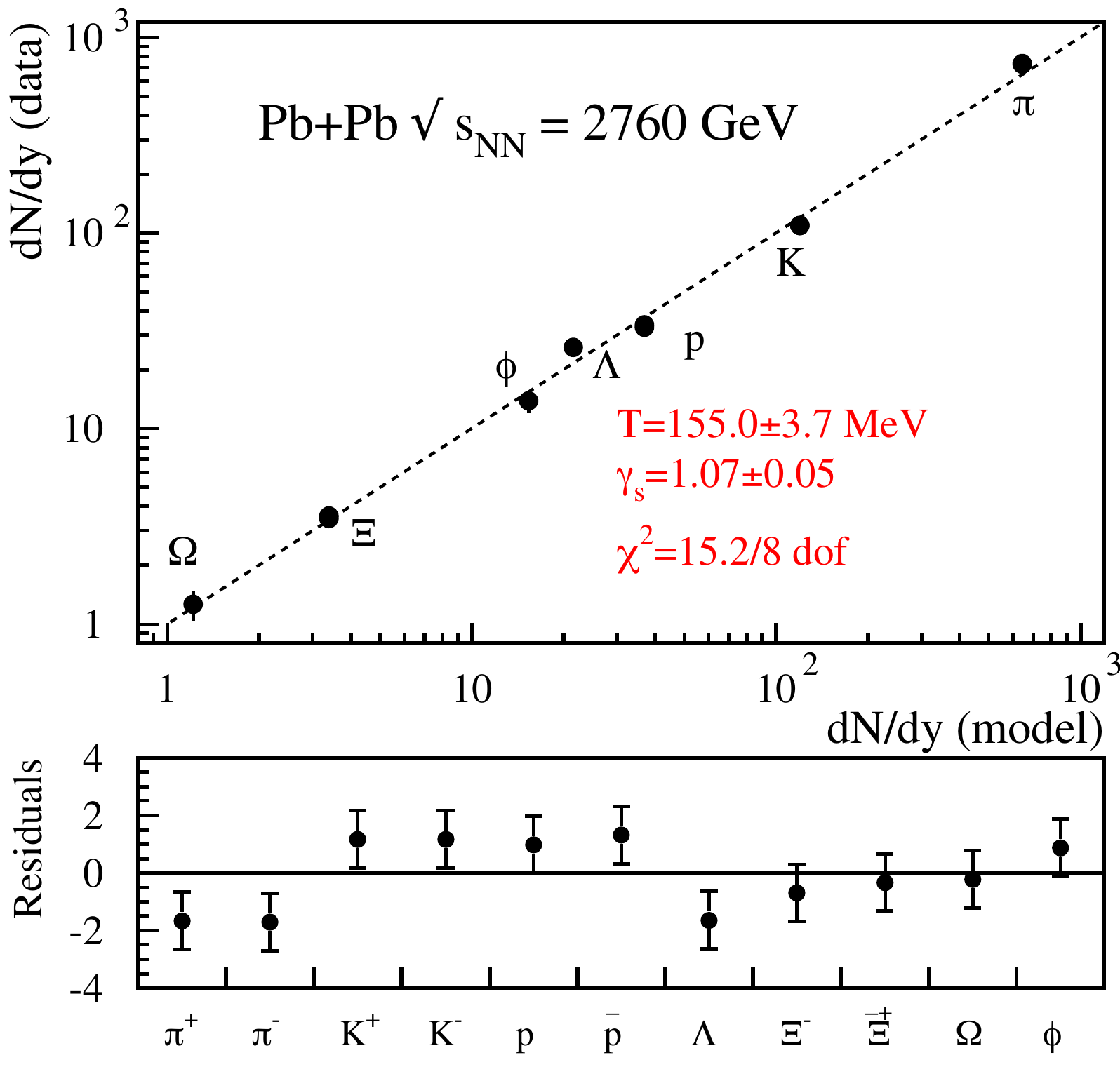}
\hspace{3pc}%
\includegraphics[width=0.38\linewidth]{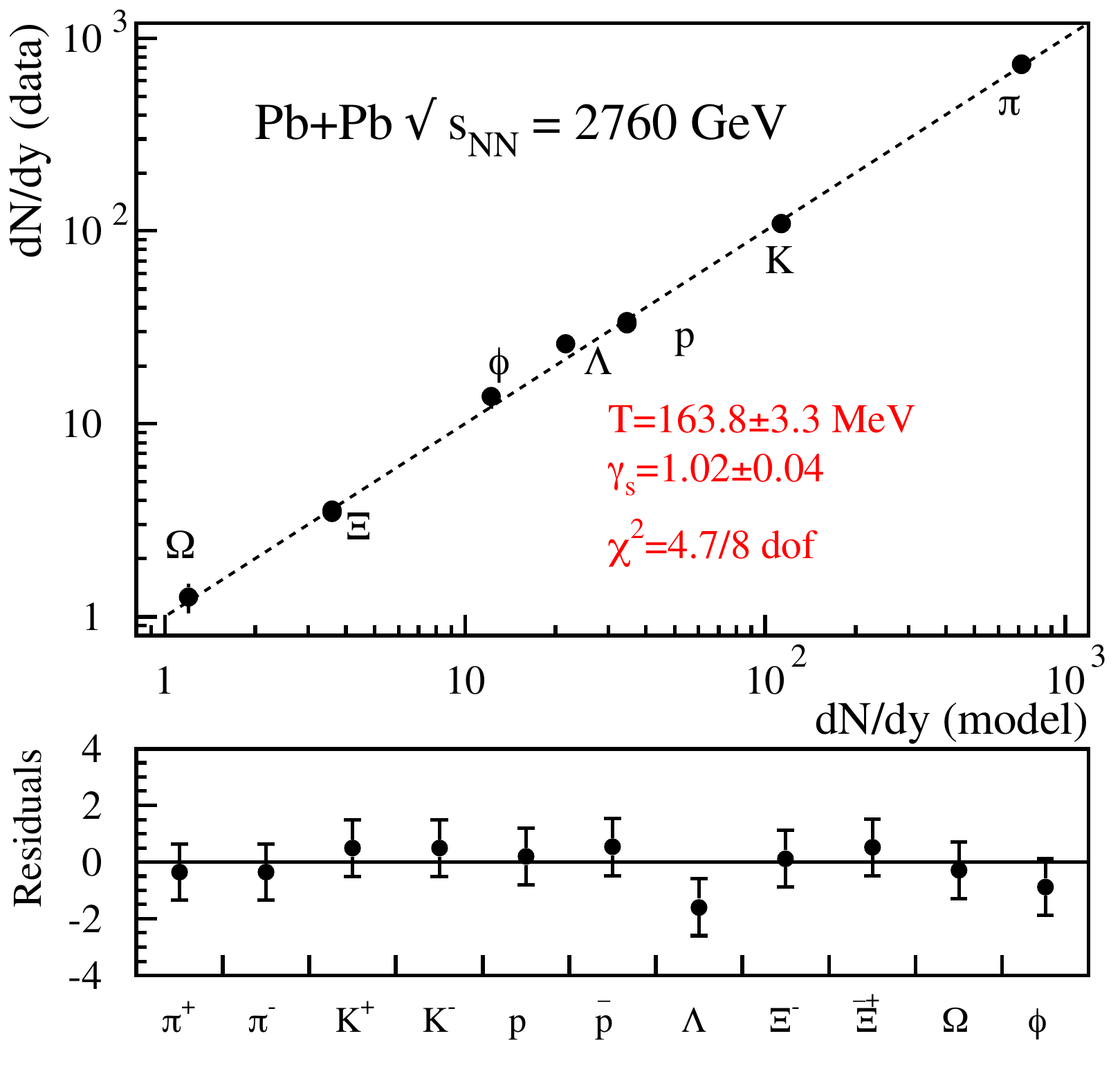}
\end{center}
\caption{Stat. model fits to central Pb+Pb data at 2.76TeV (left) and to the
      same data with UrQMD modification factors(right).}
\label{fig2}
\end{figure}
%
\section{Data analysis}
\label{Dataanalysis}
In order to illustrate the effect of the UrQMD modification factors in the SHM
analysis we show in Fig.\ref{fig2} the SHM fits to the most central ALICE Pb+Pb data at
$2.76$~TeV. The left panel shows the standard SHM procedure with $T=155\pm4$~MeV,
$\mu_{B}$ being set to zero. The fit underestimates the pions and is of moderate
overall quality, with $\chi^{2}/$dof of about two. With UrQMD modification factors(right panel) we find $T=164\pm3$~MeV, with substantially improved $\chi^{2}/$dof, of about $0.5$.
Similar observations were made at the lower energies. The deduced
temperature increases with afterburner corrections. Note that at higher $\mu_{B}$
the baryons do outnumber, by far, their anti-partners. Thus annihilation has a
much higher fractional impact on the antibaryon multiplicities, there.
%
\begin{figure}[t]
\begin{center}
\includegraphics[width=0.5\linewidth]{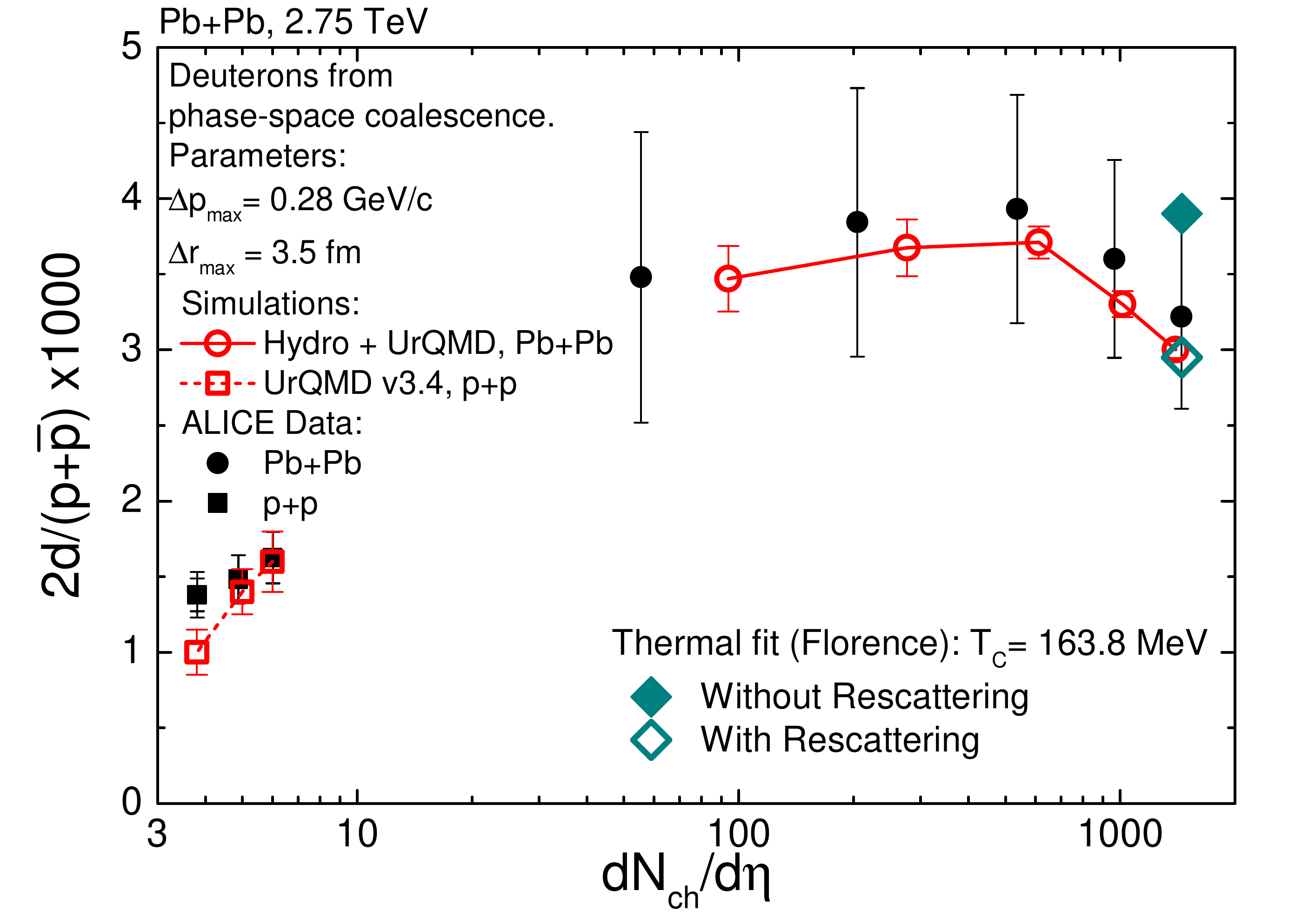}
\hspace{3pc}%
\includegraphics[width=0.35\linewidth]{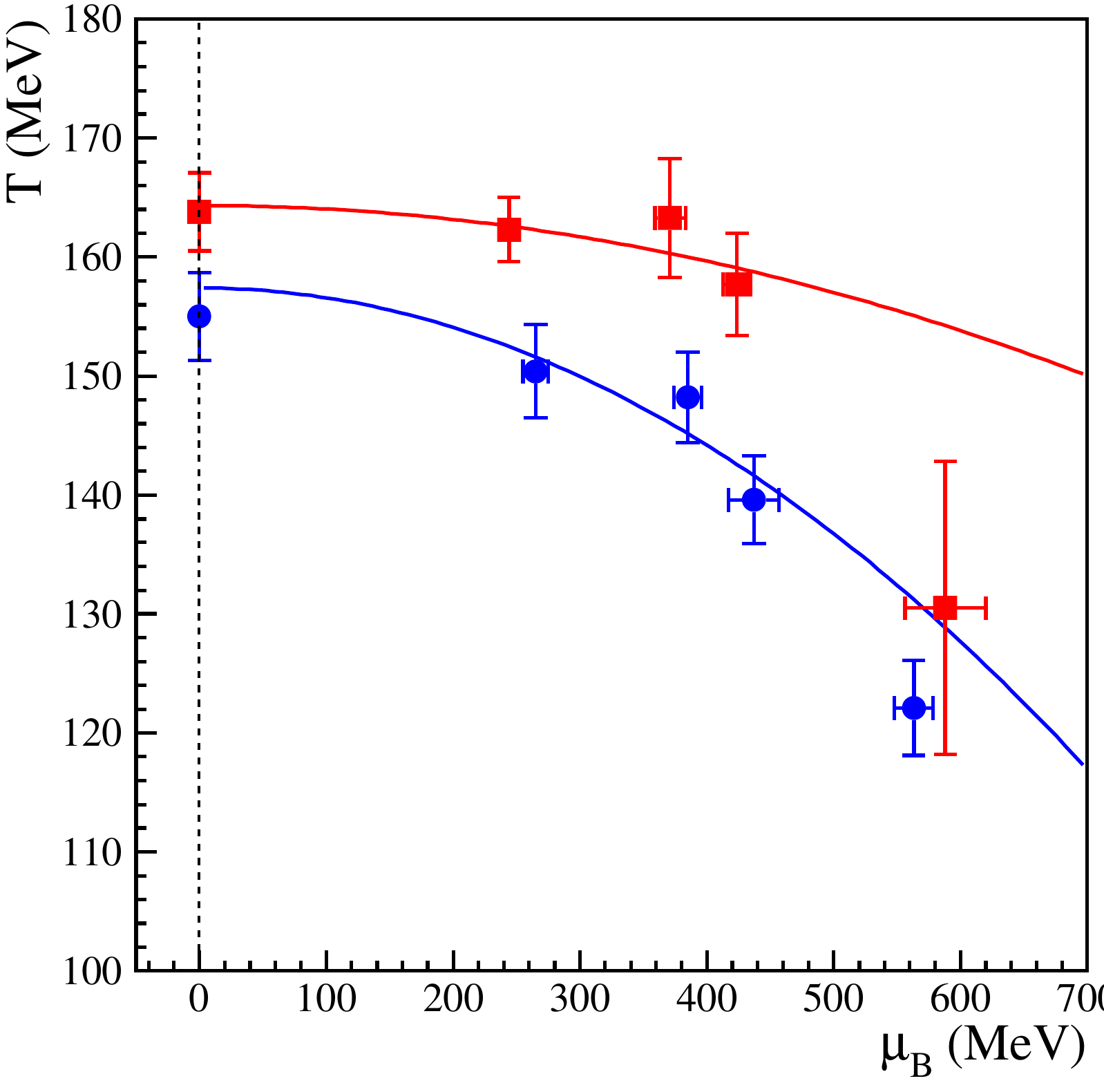}
\end{center}
\caption{Left: Fits of UrQMD with phase space coalescence, and of the statistical
      model to ALICE deuteron to proton ratio data\cite{10} for p+p and Pb+Pb.
      Right: Reconstructed hadronization points(squares) vs. standard chemical
      freeze-out points(dots) in the ($T$,$\mu_{B}$) plane, with 4 point
      quadratic fits.}
\label{fig3}
\end{figure}
%
\section{Deuteron production}
\label{Deuteronproduction}
Deuterons have been observed systematically in A+A experiments from Bevalac energies up to the LHC. They represent a tool to discuss reaction dynamics, in
particular the evolution of the entropy\cite{13}. Light nuclei can not be synthesized
at the temperatures/particle densities at hadronization and during the subsequent
hadron/resonance expansion period. A rigorous theoretical model is still lacking
but one could see their production arise from a very last stage of
final state interaction, after the end of all hadronic collision. This is expressed
in the qualitative coalescence model. It has been shown by Mrowczynski\cite{14} that the
results of the SHM should agree with the coalescence model, which corresponds to the
view that light nuclei are formed in proportion to the proton/neutron number density, and the entropy, both fixed at hadronization. We have implemented a phase
space coalescence routine at the decoupling stage of UrQMD\cite{11}. The new aspect of
this calculation consists in the simultaneous enforcement of closeness in space,
and momentum space.
We show in Fig.\ref{fig3} the ALICE data\cite{10} for 2D/(P+Pbar), equivalent to the commonly
discussed D/P ratio. The figure contains minimum bias data for p+p at various LHC energies and Pb+Pb at $2.76$~TeV, plotted vs. midrapidity $dN_{ch}/d\eta$, compared
to UrQMD calulations (version 3.4 for p+p and the hybrid version for Pb+Pb),
showing perfect overall agreement\cite{11}. The coalescence parameters (space and momentum
space radii) are kept constant, throughout. Also included is our SHM result for central
Pb+Pb collisions, obtained at $T=164$~MeV. It agrees both with the data and with
the coalescence calculation.

\section{The QCD phase diagram}
\label{QCDphasediagram}

In Fig.\ref{fig3} we show our main result, the estimated QCD pseudo-critical line in the
($T$,$\mu_{B}$) plane, as illustrated by $5$ points, for LHC, SPS and AGS
energy, respectively. The curvature is modest up to about $\mu_{B}=400$~MeV; a fit
with a quadratic ansatz reveals a curvature coefficient of $0.0048\pm0.0026$, in
qualitative agreement with recent lattice QCD calculations\cite{8,9}. The temperature at
$\mu_{B}=0$, deduced from the fit is $164\pm2$~MeV. This is higher than predicted from
lattice QCD calculations based on various different observables such as Polyakov loop or higher order fluctuations of conserved quantities\cite{15}. This might be
a consequence of the broadness of the parton-hadron transition domain, typical of a
cross-over transformation: not all observables must freeze out at the same mean
temperature. We also show in Fig.\ref{fig3} the freeze-out points obtained with the
standard SHM procedure which fall below the modification-factor corrected
results, throughout.





\begin{thebibliography}{00}


\bibitem{1}
 D.Amati and G.Veneziano, Phys.Lett. B83(1979)87
\bibitem{2}
 J.Ellis and K.Geiger, Phys.Rev. D54(1996(1967
\bibitem{3}
 F.Becattini, Z.Phys.C69(1996)485
\bibitem{4}
 F.Becattini et al., Phys.Rev.Lett. 111(2013)082302
\bibitem{5}
 F.Becattini et al., Phys.Rev.C90(2014)054907
\bibitem{6}
 F.Becattini et al., Phys.Lett. B764(2017)241
\bibitem{7}
 F.Becattini et al., Phys.Rev. C85(2012)044921
\bibitem{8}
   O.Kaczmarek et al., Phys.Rev.D83(2011)014504; 
   S.Borsanyi et al., High Energy Phys. 1208(2012)053; \\
   R.Bellwied et al, Phys.Lett. B751(2015)559
\bibitem{9}
 C.Bonati et al., Phys.Rev. D90(2014)114025 and Phys.Rev.
   D92(2015)054503
\bibitem{10}
  S.Acharya et al., ALICE Coll.,Phys.Rev.C97(2018)024615 and 
   J.Adam et al., ALICE Coll.,Phys.Rev.C93(2016)024917
\bibitem{11}
  S.Sombun et al., UrQMD Coll., arXiv:1805.11509
\bibitem{12}
  J.Stachel et al., J.Phys.Conf.Ser.509(2014)012019
\bibitem{13}
  A.Andronic et al., Phys.Lett. B697(2011)203 and refs. therein
\bibitem{14}
  S.Mrowczynski, arXiv:1607.02267
\bibitem{15}
A.Bazavov et al., Hot QCD Coll., Nucl.Phys.A931(2014)867

\end{thebibliography}



\end{document}